\documentclass{article}
\usepackage{spconf,amsmath,graphicx}

\usepackage{enumitem}
\setlist{nosep, leftmargin=14pt}

\usepackage{mwe} 
\usepackage{algorithmic}
\usepackage{algorithm}
\usepackage[outdir=./]{epstopdf}
\usepackage{subcaption}
\usepackage{kantlipsum} 
\usepackage{mwe} 
\usepackage[noadjust]{cite}

\title{MR-NOM: Multi-scale Resolution of Neuronal cells in Nissl-stained histological slices via deliberate Over-segmentation and Merging}
\name{Valentina Vadori$^{1}$ \ Jean-Marie Graïc$^{2}$ \ Livio Finos$^{3}$ \ Livio Corain$^{4}$ \ Antonella Peruffo$^2$ \ Enrico Grisan$^{1}$ }
\address{$^{1}$London South Bank University, School of Engineering, United Kingdom
\\ $^{2}$University of Padova, Dept. of Comparative Biomedicine \& Food Science, Italy
\\ $^{3}$University of Padova, Dept. of Developmental Psychology and Socialisation, Italy
\\ $^{4}$University of Padova, Dept. of Management and Engineering, Italy}
\begin{document}

\onecolumn 

\begin{center}
  \large\bfseries  
  This work has been submitted to the IEEE for possible publication. Copyright may be transferred without notice, after which this version may no longer be accessible.
\end{center}
\twocolumn 
\setcounter{page}{1} 

%
\maketitle
\begin{abstract}
In comparative neuroanatomy, the characterization of brain cytoarchitecture is critical to a better understanding of brain structure and function, as it helps to distill information on the development, evolution, and distinctive features of different populations. The automatic segmentation of individual brain cells is a primary prerequisite and yet remains challenging. A new method (MR-NOM) was developed for the instance segmentation of cells in Nissl-stained histological images of the brain. MR-NOM exploits a multi-scale approach to deliberately over-segment the cells into superpixels and subsequently merge them via a classifier based on shape, structure, and intensity features. The method was tested on images of the cerebral cortex, proving successful in dealing with cells of varying characteristics that partially touch or overlap, showing better performance than two state-of-the-art methods.

\end{abstract}

\section{Introduction}
\label{sec:intro}
Comparative neuroanatomy studies investigate anatomical changes between the brains of populations defined by factors such as sex, age, pathology, or species. The characterization of brain cytoarchitecture holds special significance in such studies, as it can provide insights into the links between the specific structure of the brain and the animal morphology, behavior, or environment \cite{AMUNTS20071061, graic2022primary,corain2020multi}. 

In a typical analysis pipeline, tissue sections (i.e., histological slices) of brain specimens are processed with Nissl stain to label neuronal cells \cite{garcia2016distinction} and are fixed for digitization as Whole Slide Images (WSI) for subsequent examination. Due to their size and complexity, WSIs are preferably processed by computerized methods, which can ensure reproducibility and speed in high-throughput pipelines, while a manual examination would be prohibitively time-consuming as well as impacted by inter- and intra-observer bias.


\begin{figure}[htb]

\begin{minipage}[b]{1.0\linewidth}
  \centering
  \centerline{\includegraphics[width=1.1\textwidth]{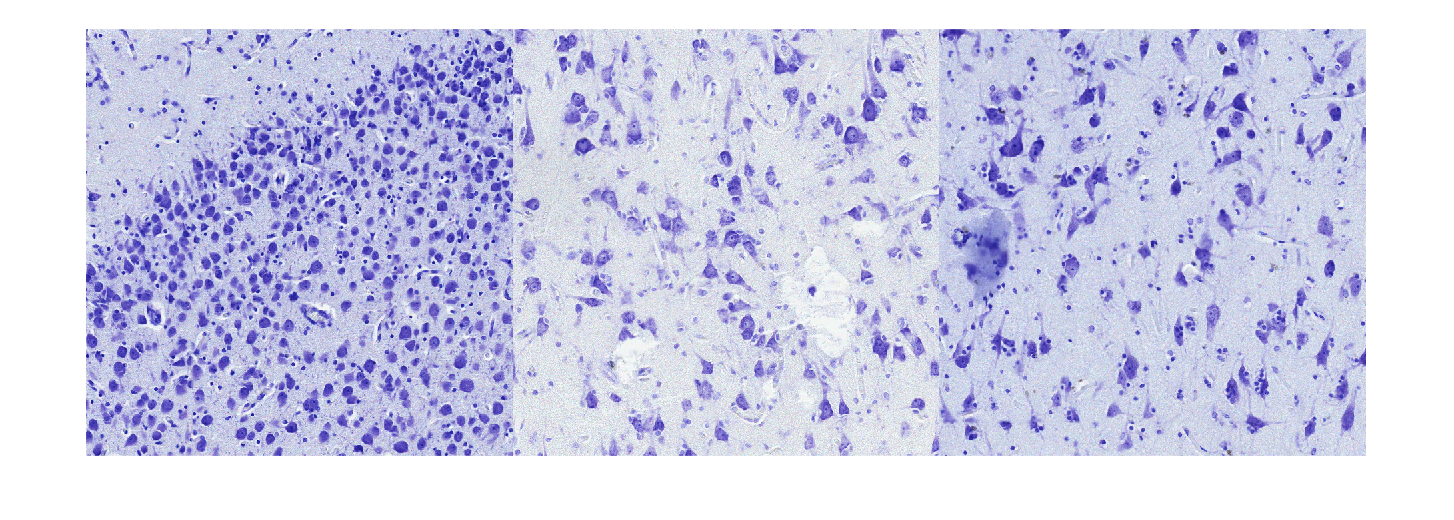}}
  \centerline{(a) Sample tiles from $40$x Nissl-stained histological slices.}\medskip
\end{minipage}
\begin{minipage}[b]{1.0\linewidth}
  \centering
  \centerline{\includegraphics[width=6cm]{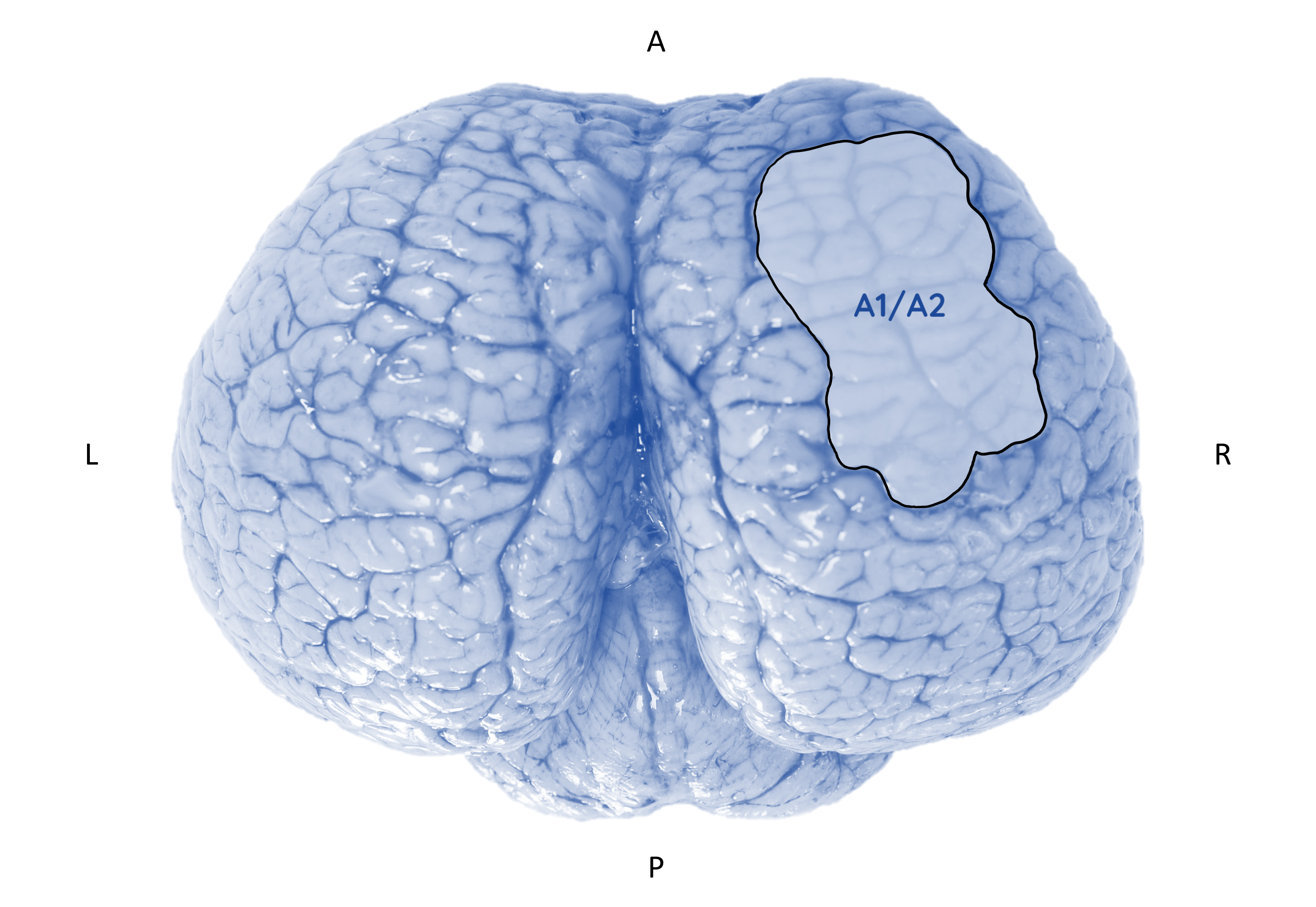}}
  \centerline{(b) Brain with highlighted primary/secondary auditory cortex.}\medskip
\end{minipage}

\caption{Samples of the auditory cortex of \textit{Tursiops truncatus}.}
\label{fig:res}
\end{figure}


A critical prerequisite in such pipelines is the challenging instance segmentation of cells. 
Fig. \ref{fig:res}a shows $3$ tiles extracted from Nissl-stained histological slices of the auditory cortex from different brain specimens of \textit{Tursiops truncatus} (Montagù, 1821), 
an example of which is shown in Fig. \ref{fig:res}b, with the highlighted area representing the primary (A1) and secondary (A2) auditory cortex. Segmentation of individual cells is complicated by their dishomogeneity in shape, texture, and size, due to the co-existence of large and small neurons, glia and endothelial cells, the presence of touching or overlapping cells with ambiguous boundaries, and background clutter.

In the wider field of digital pathology and microscopy, many segmentation methods have been proposed to segment cells/nuclei. 
The vast majority rely on a set of underlying algorithms: intensity thresholding, morphology operations, watershed transform, deformable models, clustering, graph-based approaches, and supervised classification \cite{XingRobust}. Few, however, are designed for the segmentation of cells in Nissl-stained histological slices of the brain \cite{ahrens1990image,he2015icut,grisan2018resolving}, and most are conceived for cells of uniform characteristics.

In light of the above, a new method called MR-NOM was developed to be used in an active learning fashion to facilitate the construction of ground truth annotations and to subsequently segment cells in WSIs. This method will be exploited  for the characterization of brain cytoarchitecture in comparative neuroanatomy studies, and in particular as an enabler of solid morphometric analyses aimed at objective tissue screening in the field of diseases affecting brain structure and functionality (e.g., neurodegeneration and neuroinflammation).

\begin{figure*}%
\centering
\begin{subfigure}{.29\columnwidth}
\includegraphics[width=\columnwidth]{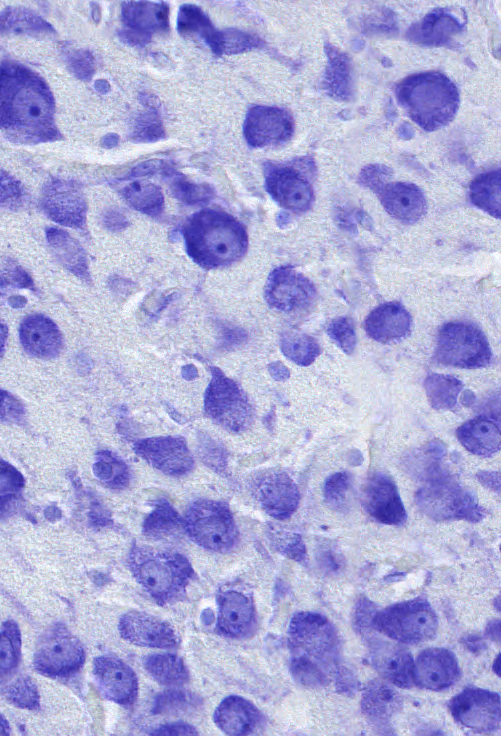}%
\caption{}%
\label{subfiga}%
\end{subfigure}\hfill%
\begin{subfigure}{.29\columnwidth}
\includegraphics[width=\columnwidth]{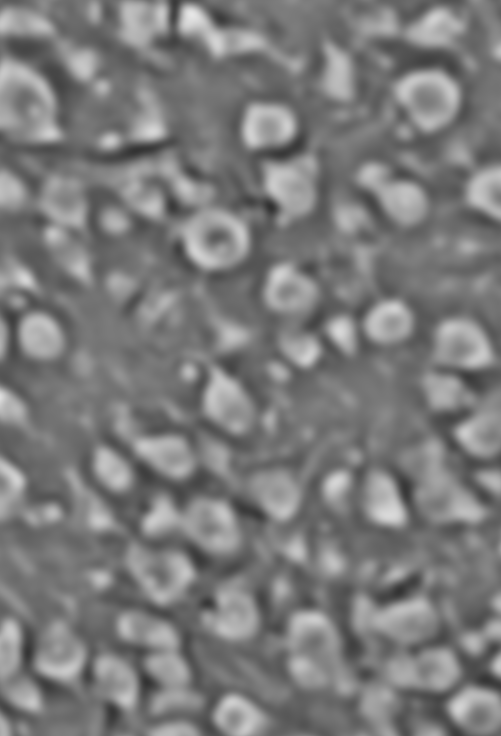}%
\caption{}%
\label{subfigb}%
\end{subfigure}\hfill%
\begin{subfigure}{.29\columnwidth}
\includegraphics[width=\columnwidth]{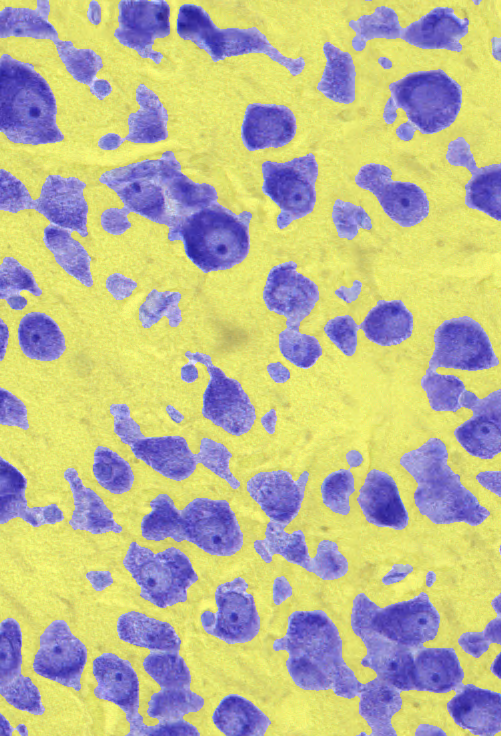}%
\caption{}%
\label{subfigc}%
\end{subfigure}\hfill%
\begin{subfigure}{.29\columnwidth}
\includegraphics[width=\columnwidth]{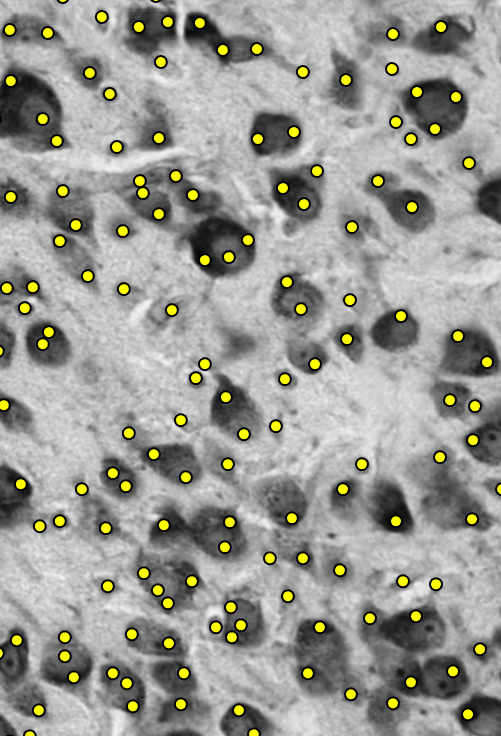}%
\caption{}%
\label{subfigd}%
\end{subfigure}\hfill%
\begin{subfigure}{.29\columnwidth}
\includegraphics[width=\columnwidth]{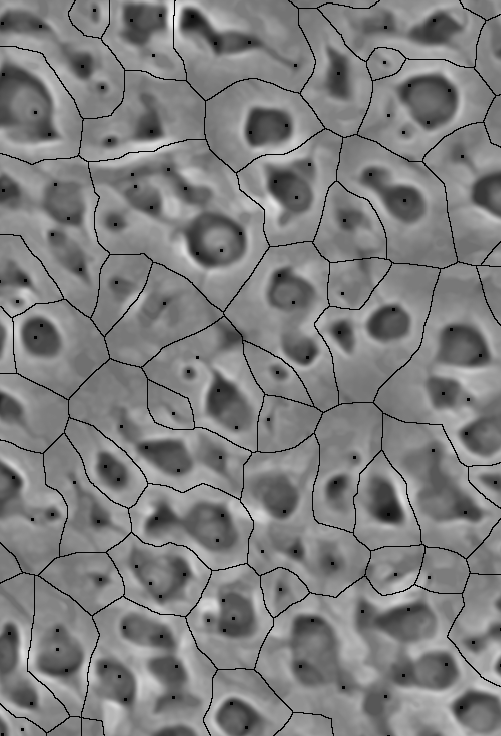}%
\caption{}%
\label{subfige}%
\end{subfigure}\hfill%
\begin{subfigure}{.29\columnwidth}
\includegraphics[width=\columnwidth]{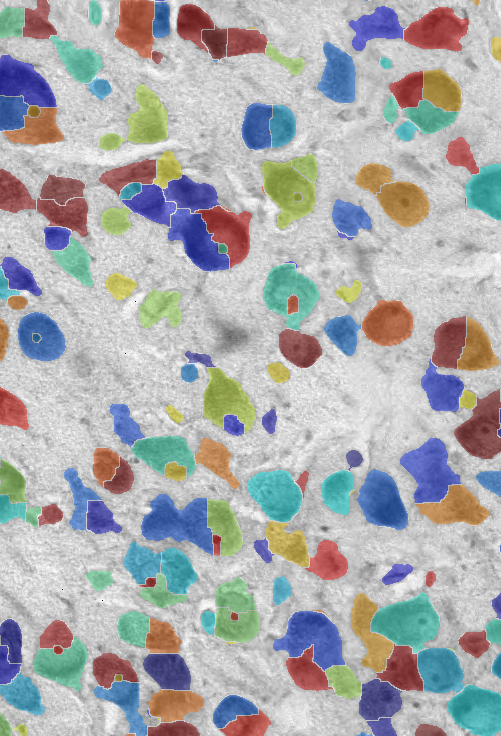}%
\caption{}%
\label{subfigf}%
\end{subfigure}\hfill%
\begin{subfigure}{.29\columnwidth}
\includegraphics[width=\columnwidth]{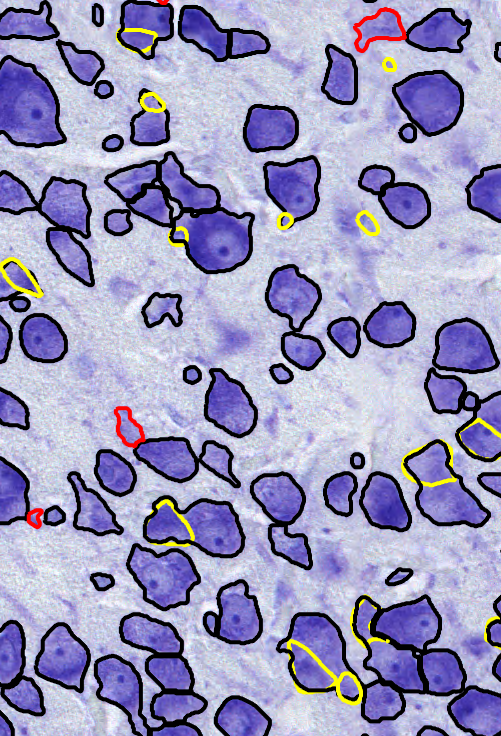}%
\caption{}%
\label{subfigg}%
\end{subfigure}\hfill%
\caption{Intermediate results of the proposed method on a sample image. (a) Original image. (b) Combined multi-scale map for foreground extraction. (c) Complement of the foreground map superimposed on the original image for illustrative purposes. (d) Detected markers in yellow. (e) Gray-scale map utilized for the watershed transform. (f) Label image returned by the watershed transform. (g) Final segmentation with true positives in black, false positives in red and false negatives in yellow.}
\label{figabc}
\end{figure*}

This paper is organized as follows: Section \ref{sec:materials} describes the dataset; Section \ref{sec:methods} details the steps of the method; Section \ref{sec:results} illustrates the results; Section \ref{sec:conclusions} draws the conclusions.

\section{Dataset}
\label{sec:materials}
Brain tissues were sampled from $20$ specimens of Tursiops truncatus archived in the Mediterranean Marine Mammals Tissue Bank (http://www.marinemammals.eu) of the University of Padova. The brains originated from stranded cetaceans with a decomposition and conservation code (DCC) of $1$ and $2$, according to the guidelines for cetacean post-mortem investigation \cite{ijsseldijk2019best}.

The images used in this study are $27$ $2048$x$2048$ tiles extracted from Nissl-stained $40$x magnification WSIs of the auditory cortex of Tursiops truncatus, also known as the bottlenose dolphin, originating from different subjects (new-born, adult, old). The tiles were annotated via QuPath \cite{bankhead2017qupath} software, leading to $13\ 986$ annotated cells. $4$ and $3$ tiles were used as validation and test set, respectively. 



\section{Methods}
\label{sec:methods}

\subsection{Pre-processing}
\label{sssec:preprocessing}

Each image was converted to grayscale and filtered with a $2$-D Gaussian smoothing filter with standard deviation (SD) of $1$. Contrast-limited adaptive histogram equalization 
was applied to enhance the contrast while avoiding noise amplification. 
The mean neuropil (area between cell bodies) intensity was standardized by applying a correcting factor as $I_1(x,y) = I_0(x,y) I_r/I_n$, where $I_r=205$ is the chosen standardized mean neuropil intensity and $I_n$ is the estimate of the mean in each equalized image $I_0(x,y)$. The latter was obtained as the mean of the two grey values at $61\%$ of the mode of the intensity histogram \cite{ahrens1990image}. 


%



The gradient map ($GM(x,y)$) was computed by convolving the standardized image $I_1(x,y)$ with a discrete $8$x$8$ filter obtained by sampling the analytical derivative of the Gaussian function with zero mean and SD of $2$. The gradient map was binarized via the triangle thresholding method \cite{zack1977automatic} 
to obtain the binary edge map ($EM(x,y)$), where connected components smaller than a threshold ($50$ pixels) were removed.
A binary Nissl-substance map ($NS(x,y)$) was obtained by applying the Otsu method \cite{otsu1979threshold} to $I_1(x,y)$. 

\subsection{Foreground extraction}
\label{sec:foregroundextraction}
A multi-scale approach based on Laplacian of Gaussian (LoG) scale-space representations was applied to foreground extraction. 
Since the application of the LoG filter at a single scale fails in detecting blobs of different sizes, a multi-scale approach is needed for detecting blobs of different (unknown) sizes. According to the scale-space theory \cite{lindeberg2013scale}, a multi-scale approach considers a set of $n$ LoG filters, with each filter $L_i, i = 1, . . . , n,$ having a different value of the standard deviation $\sigma_{i}$. The convolution of the image with each of the $n$ filters gives the set of \textit{LoG scale-space representations}:
\begin{equation}
R_i(x,y) =\sigma_{i}^\gamma \ L_i * I(x,y), \ \ \  i = 1, . . . , n
\label{scalemap}
\end{equation}
where the factor $\sigma_{i}^\gamma$ is used to normalize the response since its amplitude at blob regions decreases monotonically with increasing $\sigma_i$. Conventionally, the above set is exploited to detect local extrema and locate blobs of different scales, yielding a \textit{multi-scale LoG blob detector}. However, post-processing for blob pruning is necessary due to the large amount of overlapping blobs detected. We therefore considered the multi-scale approach presented in   \cite{kong2013generalized}, where the multiple LoG scale-space representations were summed to get a combined multi-scale map:
\begin{equation}
R(x,y) = \sum_{i=1}^{n} R_i(x,y)
\label{scalemapsum}
\end{equation}

In Nissl-stained histological slices, the objects of interest correspond to high responses in Eq. (\ref{scalemapsum}), which was exploited for foreground extraction. 
A median $3$x$3$ filter was first applied to the standardized image $I_1(x,y)$ to get $I_2(x,y)$. A combined foreground multi-scale map $R_{FG}(x,y)$ (Fig. \ref{figabc}b) was then computed as $R_{FG}(x,y) = \sum_{i=1}^{n_{FG}} R_{FG, i}(x,y)$, where $n_{FG} = 10$, $R_{FG, i}(x,y) =\sigma_{i}^{\gamma_{FG}} \ L_i * I_2(x,y)$, with $\gamma_{FG} = 1$ and $\{\sigma_{FG, i}\}_{i=1}^{n_{FG}} = \{5, 6,...,14\}$. 
$R_{FG}$ was rescaled to the intensity range $0-255$, normalized to match the mean of $I_2(x,y)$, and summed to $I_2(x,y)$ to enhance the neuronal cells. The resulting map was thresholded via the triangle method \cite{zack1977automatic} to extract the foreground, which was refined through hole-filling, morphological opening and closing, and removal of connected components smaller than $70$ pixels, H-connected, and spur pixels. 
The final foreground map $FG(x,y)$ (Fig. \ref{figabc}c) was   obtained by removing most of the poorly focused or too bright objects, using two morphological reconstructions starting from the maps $EM(x,y)$ and $NS(x,y)$ defined in Section \ref{sssec:preprocessing}.

\subsection{Marker definition}
\label{subsec:markerdefinition}
Similarly to the combined multi-scale map used to extract the foreground objects in \ref{sec:foregroundextraction}, the markers for a watershed-based over-segmentation were obtained by computing the combined multi-scale map $R_{MK}(x,y) = \sum_{i=1}^{n_{MK}} R_{MK, i}(x,y)$, where $n_{MK} = 13$, $R_{MK, i}(x,y) =\sigma_{i}^{\gamma_{MK}} \ L_i * I_1(x,y)$, with $\gamma_{MK} = 2$ and $\{\sigma_{MK, i}\}_{i=1}^{n_{MK}} = \{2, 3, ... 14\}$. The local maxima in the map were selected via the extended h-maxima transform \cite{gonzalez2018digital} with minimum height set to $8$ through hyperparameter validation. The centroids of the connected components in the binary map returned by the transform were considered as markers. The markers outside $FG(x,y)$ or too close to edges (Fig. \ref{figabc}d) were removed.
Note that, differently from \cite{kong2013generalized}, where elliptical filters with various orientations were used, only circular filters were considered here since neuronal cells have variable shapes and textures, and satisfactory results were obtained without expanding the set of filters. Furthermore, a combined multi-scale map was not only exploited for marker detection 
but also for foreground extraction, as detailed in Section \ref{sec:foregroundextraction}. 
\label{MarkerDefinition}
\subsection{Marker-controlled watershed}
\label{mkw}
The markers defined in Section \ref{subsec:markerdefinition} were used to over-segment the cells into superpixels via marker controlled-watershed \cite{gonzalez2018digital}. 
The watershed transform is typically applied to gradient maps but has also proven effective on intensity or distance transform maps and other gray-scale maps \cite{XingRobust}.

The gray-scale map used in our method integrates the combined multi-scale map defined in Section \ref{subsec:markerdefinition} with gradient information. It was defined as follows:
\begin{equation}
W(x,y)   =   R_{MK}(x,y) ^c + \alpha_1 \alpha_2 GM(x,y) 
\end{equation} 
The first term corresponds to the complement of $R_{MK}(x,y)$ as defined in Section \ref{subsec:markerdefinition}, so that objects of interest (neuronal cells) appear dark on a bright background.
The second term weighs gradient information by adding the gradient map $GM$ defined in Section \ref{sssec:preprocessing}, with $\alpha_1$ controlling the importance given to gradient cues (set to $0.15$ via hyperparameter validation). $\alpha_2=\bar{R_{MK}^c}/\bar{GM}$ is a standardization factor to make the maps comparable by matching the mean of $GM$ to the mean of $R_{MK}^c$.
Prior to any other operation, $R_{MK}$ and $GM$ were rescaled to the range $0-1$. 

$W$ was modified using morphological reconstruction to impose the markers from Section \ref{MarkerDefinition} and the \textit{SKeleton by Influence Zones} (SKIZ) of the $FG$ map as regional minima \cite{gonzalez2018digital} (Fig. \ref{figabc}e). The watershed transform was then applied to get the label image $LB(x,y)$ (Fig. \ref{figabc}f), where a different label, i.e., an integer value, is assigned to each identified region. 
All the pixels of $LB$ in the background were set to $0$.

\subsection{Supervised superpixels merging}
Ideally, each cell would correspond to a single region in the label image $LB$. However, due to cell variability, more than one marker is often associated with larger non-circular cells with diverse characteristics. These cells are represented by a set of multiple regions (or superpixels) in $LB$.
Drawing inspiration from \cite{vigueras2018corneal,stegmaier2018cell,gamarra2019split}, a classifier was trained to decide whether a pair of adjacent superpixels has to be merged.  

For every candidate merge, let $S_1$ and $S_2$ be the two superpixels to be merged, $S_{1+2}$ the resulting superpixel, $e$ the edge segment between $S_1$ and $S_2$ to be removed to create $S_{1+2}$.
The following rotation-invariant morphological, structural and intensity features were computed for $S_1$ and $S_2$ (denoted as $S$ in the descriptions): (1a) size, (2a) solidity, (3a) extent, (4a) eccentricity, (5a) circularity, (6a) axes ratio, (7a) portion of the perimeter of $S$ touching the background, (8a) ratio between the length of $e$ and the perimeter of $S$, (9a) ratio between the length of $e$ and the minor axis of the ellipse with the same second-moments as $S$, (10a) maximum, (11a) minimum, (12a) mean  intensity in $I_1(x,y)$ for pixels in $S$, (13a) SD of the intensity, (14a) intensity SD to mean ratio, (15-20a) $\mathit{1^{st}}$, $\mathit{3^{rd}}$, $\mathit{5^{th}}$, $\mathit{10^{th}}$, $\mathit{50^{th}}$ and $\mathit{75^{th}}$ intensity percentiles, (21a) maximum, (22a) minimum and (23a) mean  intensity in the gradient map $GM$ (range $0-1$) for pixels in $S$, (24a) gradient SD, (25a) gradient SD to mean ratio.

For the resulting superpixel $S_{1+2}$, some features were computed as above (all except 8-9a), and others were added: (1b) feret ratio, (2b) maximum, (3b) minimum and (4b) mean distance from the centroid of  $S_{1+2}$ to boundary points, (5b) distance SD, (6b) distance SD to mean ratio, (7b) length of $e$, (8b) ratio between the number of pixels in the intersection between the edge map $EM$ and $e$, and the length of $e$, (9b) ratio between the orientation of  $S_1$ and $S_2$, (10b) ratio between the mean intensity value in $GM$ (range $0-1$) for pixels in $e$ and the mean intensity value in $GM$ for pixels in $S_1$ and $S_2$.

The training dataset was built by processing the training images up to the marker-controlled watershed step. Pairs of adjacent superpixels in $LB$ were then considered iteratively for merging. Specifically, two iterations were performed for each connected component of the $FG$ map, typically corresponding to a single cell or a cluster of $2$ to $10$ cells. During each iteration, for each superpixel $S1$ in the connected component, the adjacent superpixels $S2$ were considered sequentially. For each candidate merge given by a pair $(S_1$, $S_2)$, $25*2+(25-2)+10=83$ features were extracted from $S_1$, $S_2$, and $S_{1+2}$, as detailed in the previous two paragraphs, and inserted into the dataset, along with the respective class ($1$ if "to be merged", $0$ otherwise), set according to ground truth. If $S1$ and $S2$ were to be merged, $S1$ was replaced by the merge $S_{1+2}$ before continuing. 
The obtained dataset was used to train a random forest classifier.

Test images were treated with the same procedure as above, with the only difference that the class of a pair of adjacent superpixels was defined by the output of the classifier.

\subsection{Post-processing}
Hole-filling, morphological opening and reconstructions were applied to the revised $LB$, followed by the removal of objects smaller than $70$ pixels and $20$ iterations of the Chan-Vese model for active contours \cite{chan2001active} to refine the cell shapes according to $I_1(x,y)$. Boundaries between touching cells were forced as defined in $LB$.
Finally, a second random forest classifier was trained on $28$ features (1-6a, 10-25a, 1-6b) of candidate cells to filter out false positive findings.

\begin{figure}[htb]
\begin{minipage}[b]{1.0\linewidth}
  \centering
  \centerline{\includegraphics[width=8.5cm]{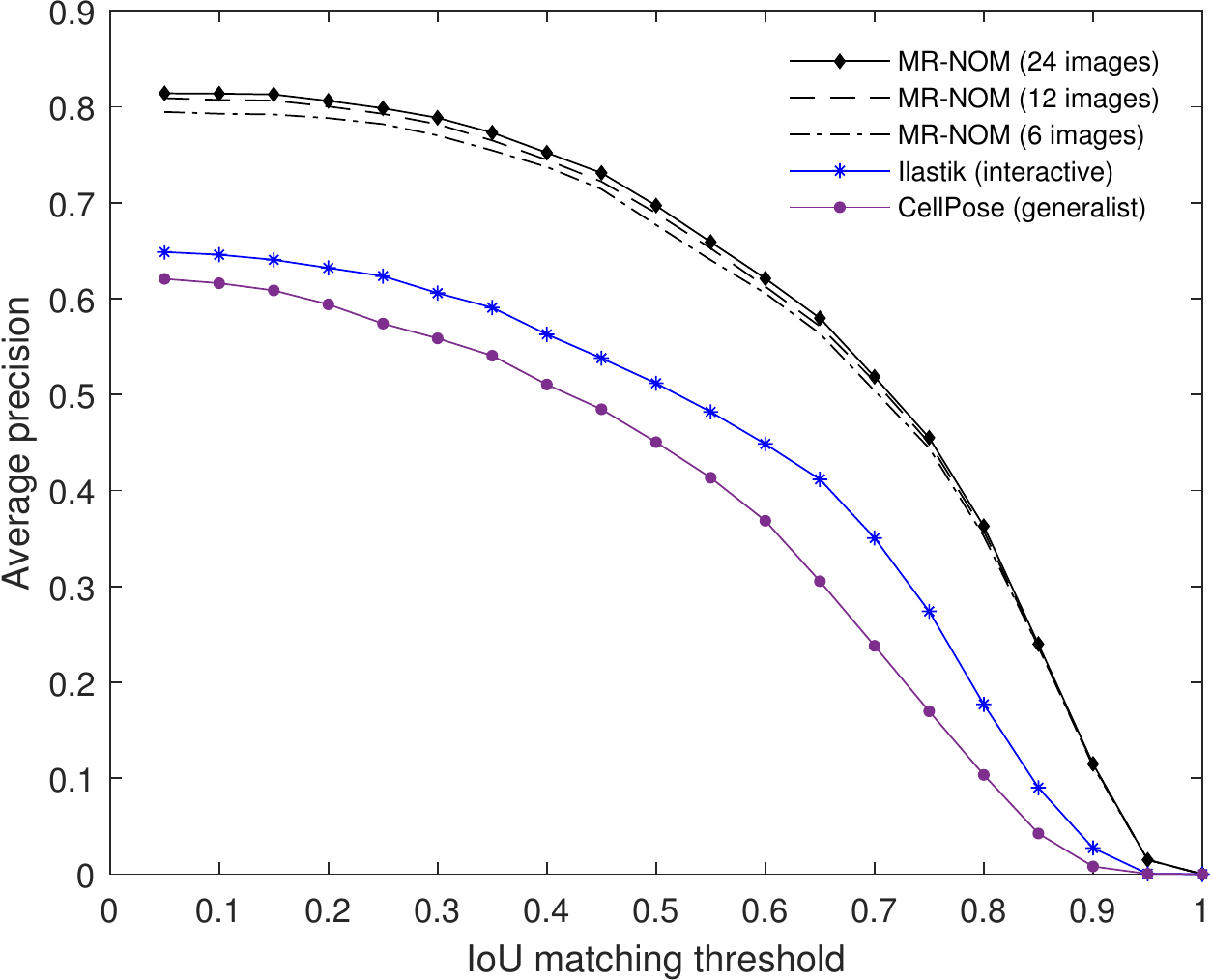}}
\end{minipage}
\caption{Segmentation performance of MR-NOM, Ilastik (interactive), and CellPose (generalist).}
\label{MRNNOMvsCellPose_AP}
\end{figure}

\section{Results}
\label{sec:results}
On test images, predictions were matched to the ground truth masks at different thresholds of matching precision based on the standard intersection over union metric (IoU). We evaluated performance with the average precision metric (AP), derived from the number of true positives (TP), false positives (FP), and false negatives (FN) as 
AP = TP/(TP+FP+FN). 

For comparison, we considered two state-of-the-art open-source solutions: Ilastik \cite{sommer2011ilastik} and the generalist CellPose model \cite{stringer2021cellpose}. Ilastik instance segmentation is attained by interactive training of a classifier to separate foreground from background, followed by hysteresis thresholding. CellPose is a deep learning-based method where a neural network is trained to predict the gradients of a topological map. These are followed via the gradient tracking process to route pixels toward the  centres of the cells and group them accordingly. The generalist model is trained on over $70\ 000$ objects. 

As shown in Fig. \ref{MRNNOMvsCellPose_AP}, MR-NOM outperformed Ilastik and CellPose at all thresholds. In particular, the AP@0.5 was $0.7$ for MR-NOM trained on $24$ images (qualitative results in Fig. \ref{figabc}g), $0.51$ for Ilastik, and $0.45$ for CellPose. It can also be observed that MR-NOM provided satisfactory results when trained on a smaller dataset. The AP@0.5 was $0.69$ and $0.68$ for MR-NOM trained on $12$ and $6$ images, respectively ($6\ 159$ and $3\ 236$ cells, respectively).
\section{Conclusions}
\label{sec:conclusions}
Few techniques have been designed for the instance segmentation of neuronal cells in Nissl-stained histological slices of the brain. We proposed a new segmentation method called MR-NOM, which exploits a multi-scale approach to deliberately over-segment the cells into superpixels to be merged via a classifier. MR-NOM dealt effectively with cells of varying characteristics that partially touch or overlap, even with a small training dataset. It was used in an active learning mode to aid the annotation process and will be exploited to segment WSIs of the auditory cortex of Tursiops truncatus. It is also expected to be adopted with suitable refinements   (e.g., more annotations and deep learning-based marker definition) to process WSIs of different species for the characterization of brain cytoarchitecture in comparative neuroanatomy studies aimed in particular at a better understanding of neurodegenerative and neuroinflammatory disorders. 
\vfill
\pagebreak

\bibliographystyle{IEEEtran}
\bibliography{refs}

\begin{thebibliography}{10}
\providecommand{\url}[1]{#1}
\csname url@samestyle\endcsname
\providecommand{\newblock}{\relax}
\providecommand{\bibinfo}[2]{#2}
\providecommand{\BIBentrySTDinterwordspacing}{\spaceskip=0pt\relax}
\providecommand{\BIBentryALTinterwordstretchfactor}{4}
\providecommand{\BIBentryALTinterwordspacing}{\spaceskip=\fontdimen2\font plus
\BIBentryALTinterwordstretchfactor\fontdimen3\font minus
  \fontdimen4\font\relax}
\providecommand{\BIBforeignlanguage}[2]{{%
\expandafter\ifx\csname l@#1\endcsname\relax
\typeout{** WARNING: IEEEtran.bst: No hyphenation pattern has been}%
\typeout{** loaded for the language `#1'. Using the pattern for}%
\typeout{** the default language instead.}%
\else
\language=\csname l@#1\endcsname
\fi
#2}}
\providecommand{\BIBdecl}{\relax}
\BIBdecl

\bibitem{AMUNTS20071061}
K.~Amunts, A.~Schleicher, and K.~Zilles, ``Cytoarchitecture of the cerebral
  cortex—more than localization,'' \emph{NeuroImage}, vol.~37, no.~4, pp.
  1061--1065, 2007.

\bibitem{graic2022primary}
J.-M. Gra{\"\i}c, A.~Peruffo, L.~Corain, L.~Finos, E.~Grisan, and B.~Cozzi,
  ``The primary visual cortex of cetartiodactyls: organization,
  cytoarchitectonics and comparison with perissodactyls and primates,''
  \emph{Brain Structure and Function}, vol. 227, no.~4, pp. 1195--1225, 2022.

\bibitem{corain2020multi}
L.~Corain, E.~Grisan, J.-M. Gra{\"\i}c, R.~Carvajal-Schiaffino, B.~Cozzi, and
  A.~Peruffo, ``Multi-aspect testing and ranking inference to quantify
  dimorphism in the cytoarchitecture of cerebellum of male, female and intersex
  individuals: a model applied to bovine brains,'' \emph{Brain Structure and
  Function}, vol. 225, no.~9, pp. 2669--2688, 2020.

\bibitem{garcia2016distinction}
M.~{\'A}. Garc{\'\i}a-Cabezas, Y.~J. John, H.~Barbas, and B.~Zikopoulos,
  ``Distinction of neurons, glia and endothelial cells in the cerebral cortex:
  an algorithm based on cytological features,'' \emph{Frontiers in
  Neuroanatomy}, vol.~10, p. 107, 2016.

\bibitem{XingRobust}
F.~Xing and L.~Yang, ``Robust nucleus/cell detection and segmentation in
  digital pathology and microscopy images: A comprehensive review,'' \emph{IEEE
  Reviews in Biomedical Engineering}, vol.~9, pp. 234--263, 2016.

\bibitem{ahrens1990image}
P.~Ahrens, A.~Schleicher, K.~Zilles, and L.~Werner, ``Image analysis of
  nissl-stained neuronal perikarya in the primary visual cortex of the rat:
  Automatic detection and segmentation of neuronal profiles with nuclei and
  nucleoli,'' \emph{Journal of Microscopy}, vol. 157, no.~3, pp. 349--365,
  1990.

\bibitem{he2015icut}
Y.~He, H.~Gong, B.~Xiong, X.~Xu, A.~Li, T.~Jiang, Q.~Sun, S.~Wang, Q.~Luo, and
  S.~Chen, ``icut: an integrative cut algorithm enables accurate segmentation
  of touching cells,'' \emph{Scientific Reports}, vol.~5, no.~1, pp. 1--17,
  2015.

\bibitem{grisan2018resolving}
E.~Grisan, J.-M. Gra{\"\i}c, L.~Corain, and A.~Peruffo, ``Resolving single
  cells in heavily clustered nissl-stained images for the analysis of brain
  cytoarchitecture,'' in \emph{2018 IEEE 15th International Symposium on
  Biomedical Imaging (ISBI 2018)}, 2018, pp. 427--430.

\bibitem{ijsseldijk2019best}
L.~L. IJsseldijk, A.~C. Brownlow, and S.~Mazzariol, ``Best practice on cetacean
  post mortem investigation and tissue sampling,'' \emph{Jt. ACCOBAMS ASCOBANS
  Doc}, pp. 1--73, 2019.

\bibitem{bankhead2017qupath}
P.~Bankhead, M.~B. Loughrey, J.~A. Fern{\'a}ndez, Y.~Dombrowski, D.~G. McArt,
  P.~D. Dunne, S.~McQuaid, R.~T. Gray, L.~J. Murray, H.~G. Coleman
  \emph{et~al.}, ``Qupath: open source software for digital pathology image
  analysis,'' \emph{Scientific Reports}, vol.~7, no.~1, pp. 1--7, 2017.

\bibitem{zack1977automatic}
G.~W. Zack, W.~E. Rogers, and S.~A. Latt, ``Automatic measurement of sister
  chromatid exchange frequency,'' \emph{Journal of Histochemistry \&
  Cytochemistry}, vol.~25, no.~7, pp. 741--753, 1977.

\bibitem{otsu1979threshold}
N.~Otsu, ``A threshold selection method from gray-level histograms,''
  \emph{IEEE Transactions on Systems, Man, and Cybernetics}, vol.~9, no.~1, pp.
  62--66, 1979.

\bibitem{lindeberg2013scale}
T.~Lindeberg, \emph{Scale-Space Theory in Computer Vision}.\hskip 1em plus
  0.5em minus 0.4em\relax Springer US, 1993.

\bibitem{kong2013generalized}
H.~Kong, H.~C. Akakin, and S.~E. Sarma, ``A generalized laplacian of gaussian
  filter for blob detection and its applications,'' \emph{IEEE Transactions on
  Cybernetics}, vol.~43, no.~6, pp. 1719--1733, 2013.

\bibitem{gonzalez2018digital}
R.~C. Gonzalez and R.~E. Woods, \emph{Digital Image Processing, Global Edition,
  4th edition}.\hskip 1em plus 0.5em minus 0.4em\relax Pearson, 2018.

\bibitem{vigueras2018corneal}
J.~P. Vigueras-Guill{\'e}n, E.-R. Andrinopoulou, A.~Engel, H.~G. Lemij, J.~van
  Rooij, K.~A. Vermeer, and L.~J. van Vliet, ``Corneal endothelial cell
  segmentation by classifier-driven merging of oversegmented images,''
  \emph{IEEE Transactions on Medical Imaging}, vol.~37, no.~10, pp. 2278--2289,
  2018.

\bibitem{stegmaier2018cell}
J.~Stegmaier, T.~V. Spina, A.~X. Falcao, A.~Bartschat, R.~Mikut, E.~Meyerowitz,
  and A.~Cunha, ``Cell segmentation in 3d confocal images using supervoxel
  merge-forests with cnn-based hypothesis selection,'' in \emph{2018 IEEE 15th
  International Symposium on Biomedical Imaging (ISBI 2018)}, 2018, pp.
  382--386.

\bibitem{gamarra2019split}
M.~Gamarra, E.~Zurek, H.~J. Escalante, L.~Hurtado, and H.~San-Juan-Vergara,
  ``Split and merge watershed: A two-step method for cell segmentation in
  fluorescence microscopy images,'' \emph{Biomedical Signal Processing and
  Control}, vol.~53, p. 101575, 2019.

\bibitem{chan2001active}
T.~F. Chan and L.~A. Vese, ``Active contours without edges,'' \emph{IEEE
  Transactions on Image Processing}, vol.~10, no.~2, pp. 266--277, 2001.

\bibitem{sommer2011ilastik}
C.~Sommer, C.~Straehle, U.~Köthe, and F.~A. Hamprecht, ``Ilastik: Interactive
  learning and segmentation toolkit,'' in \emph{2011 IEEE International
  Symposium on Biomedical Imaging: From Nano to Macro}, 2011, pp. 230--233.

\bibitem{stringer2021cellpose}
C.~Stringer, T.~Wang, M.~Michaelos, and M.~Pachitariu, ``Cellpose: a generalist
  algorithm for cellular segmentation,'' \emph{Nature Methods}, vol.~18, no.~1,
  pp. 100--106, 2021.

\end{thebibliography}

\end{document}